\documentclass[prd,aps,tightenlines,a4paper,12pt]{revtex4}
\usepackage{graphicx}
\newcommand{\ve}[1]{\mbox{\boldmath$#1$}}
\begin{document}
\title{Light deflection in binary stars}
\author{Sven Zschocke}
\affiliation{
Lohrmann Observatory, Dresden Technical University,\\
Mommsen Str. 13, D-01062 Dresden, Germany\\
}

\begin{abstract}
The light deflection of one component of a binary system due to the 
gravitational field of the other component is investigated. While this 
relativistic effect has not been observed thus far, the question arises that  
whether this effect becomes detectable in view of todays high-precision astrometry 
which soon will reach the microarcsecond level of accuracy. 
The effect is studied and its observability is investigated.
It turns out, that in total there are about $10^3$ binaries having orbital 
parameters such that the light deflection amounts to be at least $1$ 
microarcsecond. Two stringent criteria for the orbital parameters are presented, by 
means of which one can easily determine the maximal value of light deflection 
effect for a given binary system. It is found, that for relevant binaries their orbital parameters 
must take rather extreme values in order to have a light deflection of the order of a few microarcseconds. 
Only in a very few and rather extreme binary systems the light deflection 
effect might be detectable by todays astrometry, but their existence is highly 
improbable. Thus, the detection of this subtle effect of relativity still remains 
a challenge for future astrometric missions.
\end{abstract}

\maketitle

\newpage

\tableofcontents

\newpage


\section{Introduction}\label{Section0}

Astrometric space missions, especially the ESA (European Space Agency) cornerstone mission Gaia, 
see e.g. {\it Perryman, et al.} \cite{Gaia_Overview}, are in preparation to attain microarcsecond 
($\mu{\rm as}$) level of accuracy in absolute positional measurements of stars and other celestial objects. This 
unprecedented accuracy of astrometric observations makes it necessary to account for many subtle effects which were 
totally negligible before. A practical model for astrometric observations with an accuracy of $1\,\mu{\rm as}$ has been 
formulated by {\it Klioner} \cite{Klioner1}, where the influence of the gravitational fields inside the solar system were 
taken into account. Furthermore, a number of additional effects potentially observable at this level of accuracy due to 
various gravitational fields generated outside of the solar system were also briefly discussed in that investigation. One 
of them is the gravitational light deflection of one companion of a binary system in the gravitational field of the other 
companion (without loss of generality, throughout the investigation the light deflection of component B at component A is 
considered, hence, component A is considered to be the massive body, while component B is the light-source). 
This effect would change the apparent position of one component of a binary system. 
While it is clear that this effect is relatively small and even at the level of $1\,\mu{\rm as}$ observable 
only for edge-on binary systems, the huge amount of binaries generate some hope that there are relevant systems where 
this light deflection effect becomes detectable. 
For instance, Gaia will observe $10^9$ stars brighter than $20^{\rm th}$ apparent magnitude. Detailed numerical simulations 
predict the detection of about $10^8$ (resolved, astrometric, eclipsing, spectroscopic) binary systems by Gaia mission, see 
{\it Zwitter {\rm \&} Munari} \cite{GAIA_Binary1}, which is a considerable increase compared to the $10^5$ binary systems 
known so far, see "{\it Washington Double Star Catalog}" \cite{Washington_Double_Star}.

Let us consider this argument a bit more quantitatively. For a simple estimate of the expected order of magnitude in 
light deflection, the classical lens equation (in the form given by Eq.(67) in {\it Fritelli, et al.} 
\cite{Fritelli_Kling_Newmann}, Eq.~(24) in {\it Bozza} \cite{Bozza} or Eq.~(23) in {\it Zschocke} 
\cite{Article_Generalized_Lens_Equation}) is applied. 
In terms of orbital elements of a binary system it can be written as follows:
\begin{eqnarray}
\varphi = \frac{1}{2} \left(\sqrt{\frac{A^2}{r^2}\,\cos^2 i + 16\,\frac{m}{r}\,\frac{A}{r}\,\sin i}
- \frac{A}{r}\,\left|\cos i\right|\right).
\label{classical_lens_1}
\end{eqnarray}

\noindent
Here, $\varphi$ is the light deflectin angle, $i$ is the inclination, $A$ is the semi-major axis and $r$ 
is the distance of center-of-mass of the binary system from the observer, and the Schwarzschild radius 
is $\displaystyle m= \frac{G\,M}{c^2}$, where $G$ is the gravitational 
constant and $c$ is the speed of light, and $M$ is the stellar mass of component A of the binary system. 
From (\ref{classical_lens_1}) one obtains the maximal possible light deflection for edge-on binaries,
i.e. attained for the case when the inclination is exactly $90^{\circ}$:
\begin{eqnarray}
\varphi &\le& 200\,\mu{\rm as}\;\sqrt{\frac{M}{M_{\odot}}\;\frac{A}{\rm AU}}\;\frac{\rm pc}{r}\;.
\label{classical_lens_2}
\end{eqnarray}

\noindent
Here, ${\rm AU} = 1.496 \times 10^{11}\;{\rm m}$ is the astronomical unit, 
${\rm pc} = 3.086 \times 10^{16}\;{\rm m}$ stands for parallax of one arcsecond, and $M_{\odot}$ is the solar mass. 
According to this formula, the choice of moderate values like $M=M_{\odot}$ and $A \sim 100\,{\rm AU}$ would result into 
significant light deflection effect on microarcsecond level even at large distances of about $r \sim 100 \,{\rm pc}$. 
A meaningful value for the density of stars in the solar neighborhood, $0.025 \,{\rm binaries}\;{\rm pc}^{-3}$, implies 
already $10^5$ binaries inside a sphere of $r = 100\,{\rm pc}$. Thus, one might conclude among them there are 
a few relevant edge-on binary systems.  

However, in order to estimate quantitatively the number of relevant systems, one needs to know the probability 
for the occurrence of such edge-on binaries which have a given light deflection depending on their orbital parameters like 
inclination, mass, distance and semi-major axis. Such a relation between a given light deflection and orbital parameters is 
given by a so-called inclination formula. Since the inclinations of binary systems are of course randomly distributed,  
it is meaningful to resolve such an inclination formula in terms of inclination.  
As it has been shown by {\it Klioner, et al.} \cite{inclination_formula} (see Appendix \ref{Appendix_KMS}
for some basic steps), such an inclination formula can be obtained 
by means of the analytical solution of light deflection in standard post-Newtonian approach, and is given as follows: 
\begin{eqnarray}
\left|\frac{\pi}{2} - i \right|_{\rm KMS} &\le& 2\,\arctan \left(0.0197\;\frac{M}{M_{\odot}}\;
\frac{\mu{\rm as}}{\varphi}\;\frac{\rm pc}{r}\right).
\label{KMS_D}
\end{eqnarray}

\noindent
For a better illustration of the inclination formula, relation (\ref{KMS_D}) is rewritten in terms of angular degrees 
instead of radians:
\begin{eqnarray}
\left| \,90^{\circ} - i\,\right| &\le& 2.25^{\circ}\;\frac{M}{M_{\odot}}\;\frac{\mu{\rm as}}{\varphi}\;\frac{\rm pc}{r}\;,
\label{condition_0}
\end{eqnarray}

\noindent
where also $\arctan x = x + {\cal O} (x^3)$ has been used. According to this relation, the inclination $i$ of a binary 
system with stellar mass $M$ and at distance $r$ must not deviate from the edge-on value $90^{\circ}$ 
too much in order to have a given light deflection $\varphi$.
For example, for a hypothetical binary star with $M=M_{\odot}$ situated at a distance of $r=10\,{\rm pc}$
the light deflection effect attains $1\,\mu{\rm as}$ only if $\left| \,90^{\circ} - i\,\right| < 0.225^{\circ}$,
which means that the probability to observe this binary at a favorable inclination is only about $0.2\%$.

But even this estimation is still much too optimistic. 
As a concrete example of todays high-precision astrometry, let us consider one important parameter about the astrometric 
accuracy of the Gaia mission: the accuracy of one individual positional measurement in the most ideal case (bright star, i.e. $10^{\rm th}$ magnitude) 
amounts to be $25\,\mu{\rm as}$, which implies $\varphi \ge 25\,\mu{\rm as}$. Furthermore, inside a sphere of $10\,{\rm pc}$ 
around the Sun almost every star and binary system is known already by the data of  
{\it Research Consortium on Nearby Stars (RECONS)} \cite{Recons}. 
Since inside that sphere there is no binary system having a light deflection on microarcsecond level, 
one has to take at least $r > 10\,{\rm pc}$. By taking into account these both remarks, one obtains  
$\displaystyle \frac{\mu{\rm as}}{\varphi}\;\frac{\rm pc}{r} = \frac{1}{250}$ in relation (\ref{condition_0}). Therefore, 
even in the best case one has to conclude $\left| \,90^{\circ} - i\,\right| \le 0.01^{\circ}\;M/M_{\odot}$, that means 
(for $M=M_{\odot}$), the probability to observe such a binary at a favorable inclination is practically only 
about $0.01\%$. And the binaries must be, in fact, almost edge-on in order to have a light deflection which can be detected 
by todays astrometry. Accordingly, while relation (\ref{classical_lens_2}) triggers the hope about the existence of 
many relevant binary systems, from relation (\ref{condition_0}) one concludes that the number of relevant binary systems 
is considerably reduced. 

An estimation of the total amount of binaries depends on many different parameters, like mass, semi-major axis, 
inclination and distances of the binaries. Therefore, a simple estimation is not so straightforward as one 
might believe. Moreover, the relation (\ref{classical_lens_2}) has been obtained with the aid of classical 
lens equation, while relation (\ref{condition_0}) has been obtain by means of standard post-Newtonian approach. 
These both approaches have different regions of validity. However, a rigorous treatment of the problem of light deflection 
in binary systems implies the need of an analytical formula which is valid for such kind of extreme astrometric 
configurations like the binary systems are. Recently, a generalized lens equation has been derived by {\it Zschocke} 
\cite{Article_Generalized_Lens_Equation}, which allows to determine the light deflection of binary systems on 
microarcsecond level. One aim of this study is, therefore, to reobtain the criteria (\ref{classical_lens_2}) and
(\ref{condition_0}) as stringent conditions from one and the same approach, i.e. with the aid of generalized lens equation. 
This is possible, because in the corresponding limits the generalized lens equation agrees with the classical lens equation 
and the standard post-Newtonian solution. 
Another aim is, to derive an inclination formula like (\ref{KMS_D}) for binary systems which allows to determine the 
needed inclination for a given light deflection angle and as a function of the orbital parameters. For that one has to 
take into account the distribution of stellar masses and the distribution of semi-major axes in binary systems.
Finally, the aim of this study is, to determine the total number of relevant binaries having a light deflection on 
microarcsecond level, and to investigate the possibility to detect this effect of light deflection 
by todays high precision astrometry. 

The article is organized as follows: 
In Section \ref{Section1} some basics about orbital elements of binary systems
are given. The generalized lens equation and the inclination formula are presented in Section \ref{Section2}.
In Section \ref{Condition1} two stringent conditions on the orbital elements of binary systems 
(astrometric, spectroscopic, eclipsing and resolved binaries) are presented, which allow to determine 
whether or not the binary system will have a light deflection of a given magnitude. The total number of binaries which have 
a given light deflection for an infinite time of observation is estimated in Section \ref{Section3a}, while the more 
practical case of a finite time of observation is considered in Section \ref{Section3b}. 
The special case of resolved binaries is considered in Section \ref{Condition2}. For that, the specific instrumentation of 
Gaia mission is considered in some detail as the most modern astrometric mission with todays highest possible accuracy. 
A Summary is given in Section \ref{Summary}.

\section{Orbital elements of a binary system}\label{Section1}

\begin{figure}
\centering
\includegraphics[width=12.0cm]{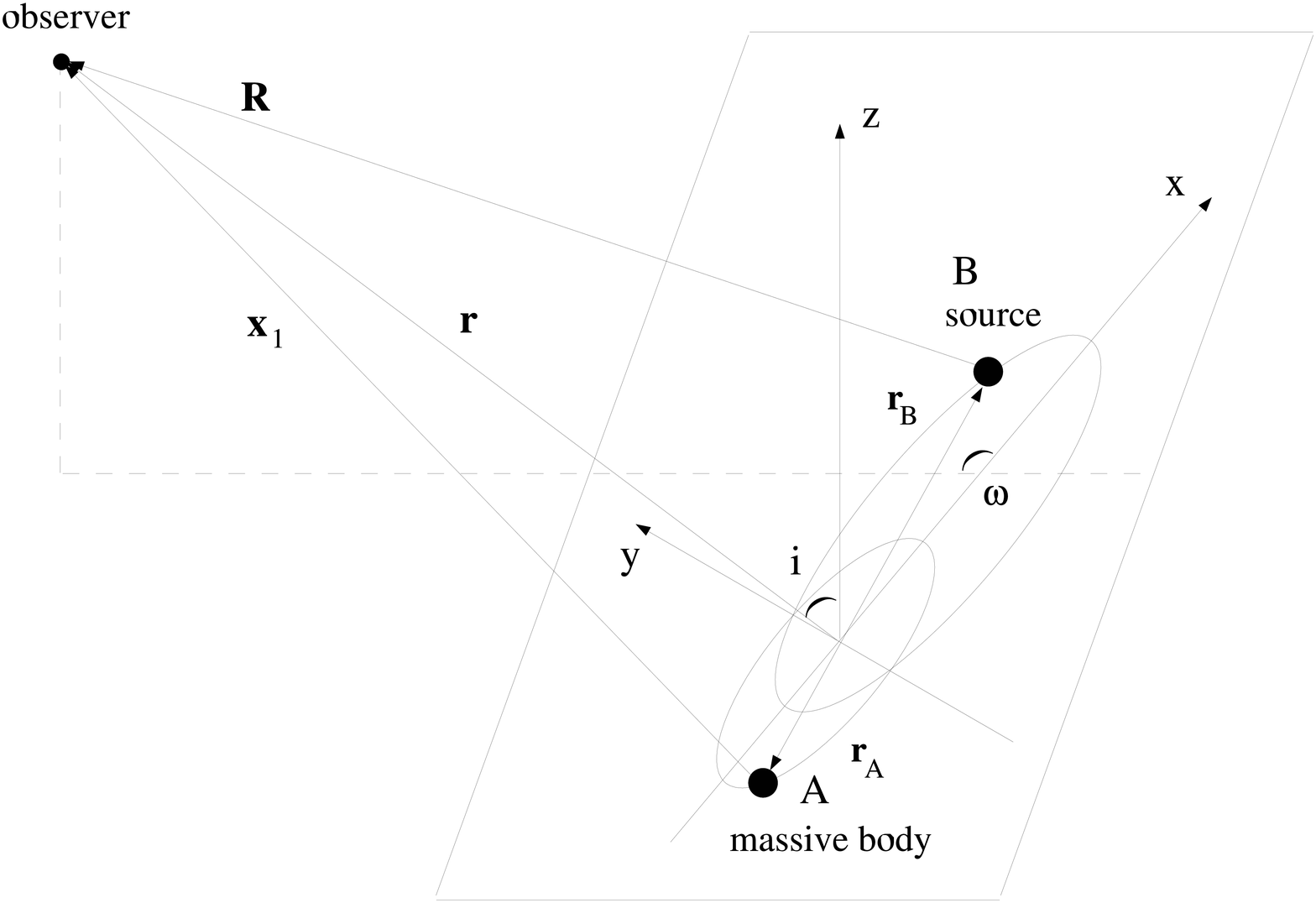}
\caption{Seven orbital elements which define the orbit of a binary system: distance vector $\ve{r}$, semi-major axis
$A$, inclination $i$, eccentricity $e$, eccentric anomaly $E$, periapsis $\omega$, and mass ration $M_A/M_B$.
The orbit of the binary system spans the $(x,y)$-plane and the $z$-axis is perpendicular
to the orbital plane. The $x$-axis is oriented along the semi-major axis of the orbit of the binary system, while
the $y$-axis is perpendicular to the $x$-axis. The vector $\ve{r}$ is directed from the center-of-mass (CMS) of the
binary system, see Eq.~(\ref{inclination_3}), to the observer. The center of spherical coordinate system is located at
the CMS of binary system, i.e. $\ve{r}_{\rm CMS} = \ve{0}$. The inclination $0 \le i \le \pi$ is the angle
between $\ve{r}$ and $z$-axis; $i = \pi/2$ is called edge-on and $i > \pi/2$
corresponds to retrograde orbit. The dotted line indicates the projection of $\ve{r}$ onto orbital $(x,y)$-plane, i.e.
z-component of $\ve{r}$ equals zero. The angle between this projection and $x$-axis is called argument of periapsis
$0 \le \omega \le \pi$. The orbital elements semi-major axis $A$, eccentricity $0 \le {\rm e} \le 1$ and mass ratio
$M_A/M_B$ govern uniquely the geometric shape of both ellipses. The eccentric anomaly $0 \le E \le 2\,\pi$
(not plotted here), is defined in Eq.~(\ref{eccentric_anomaly_5}) of Appendix \ref{AppendixB} and determines the
actual position of the bodies A and B on their orbit.}
\label{FIG: Orbital_Elements}
\end{figure}

Consider a binary system, component $A$ with mass $M_A$ at coordinate $\ve{r}_A$ and component $B$ with mass $M_B$ 
at coordinate $\ve{r}_B$. In order to express the light deflection effect in terms of orbital elements, spherical 
coordinates are introduced, illustrated by FIG.~\ref{FIG: Orbital_Elements}.

The center of coordinate system is located at the CMS, i.e.
\begin{eqnarray}
\ve{r}_{\rm CMS} &=& \frac{1}{M_A + M_B} \left( M_A \;\ve{r}_A + M_B \;\ve{r}_B \right)\,.
\label{inclination_3}
\end{eqnarray}

\noindent
Thus, the vector $\ve{r}$, which points from CMS to the observer, is given by
\begin{equation}
\ve{r} = \left( \begin{array}[c]{c}
\displaystyle
r\,\cos \omega\;\sin i \\
\nonumber\\
\displaystyle
r\,\sin \omega\;\sin i \\
\nonumber\\
\displaystyle
r\,\cos i
\end{array}\right)\,,
\label{inclination_5}
\end{equation}

\noindent
where $r = \left|\,\ve{r}\,\right|$, the argument of periapsis is denoted by $\omega$, and $i$ is the inclination,
see FIG.~\ref{FIG: Orbital_Elements}. The solution of equation of motion yields for vectors $\ve{r}_A$ and $\ve{r}_B$
the expression given by Eqs.~(\ref{appendixB_56}) - (\ref{appendixB_62}).
The vector $\ve{x}_1$ points from the mass center of massive body to the observer, and vector $\ve{x}_0$
points from the mass center of massive body to the source, see also FIG.~\ref{FIG: Orbital_Elements}. The coordinates
of these vectors can be expressed by the orbital elements of the binary star as follows:
\begin{equation}
\ve{x}_1 = \ve{r} - \ve{r}_{\rm A} = \left( \begin{array}[c]{c}
\displaystyle
r\,\cos \omega\;\sin i - \frac{A\, \left(\cos E - {\rm e} \right)}{1 + \frac{\displaystyle M_A}{\displaystyle M_B}}\\
\nonumber\\
\displaystyle
r\,\sin \omega\;\sin i - \frac{A\,\sqrt{1 - {\rm e}^2}\sin E}{1 + \frac{\displaystyle M_A}{\displaystyle M_B}}\\
\nonumber\\
\displaystyle
r\,\cos i
\end{array}\right)\,,
\label{inclination_10}
\end{equation}

\begin{equation}
\ve{x}_0 = \ve{r}_{\rm B} - \ve{r}_{\rm A} = - \left( \begin{array}[c]{c}
A \; \left(\cos E - {\rm e} \right)  \\
\nonumber\\
\displaystyle
A\; \sqrt{1 - {\rm e}^2} \sin E \\
\nonumber\\
\displaystyle
0
\end{array}\right)\,.
\label{inclination_15}
\end{equation}

\noindent
Here, $A$ is the semi-major axis, $e$ is the eccentricity, and $E$ is the eccentric anomaly, see Appendix \ref{AppendixB}.
The vectors (\ref{inclination_10}) and (\ref{inclination_15}) will be used to express the light deflection
in terms of orbital elements of the binary system.

\section{Inclination formula from generalized lens equation}\label{Section2}

\begin{figure}[h!]
\centering
\includegraphics[width=14.0cm]{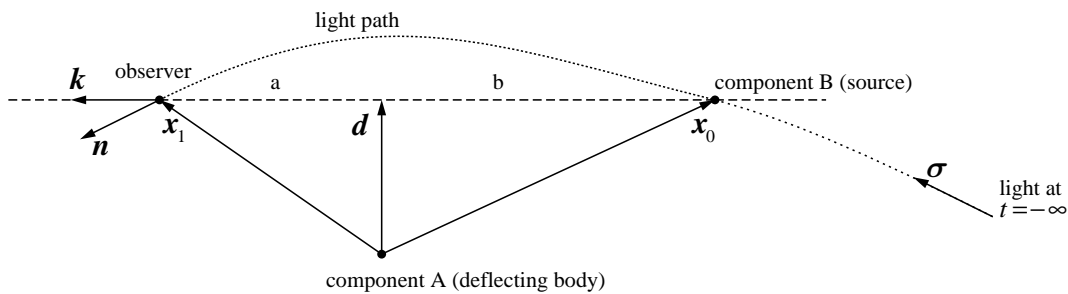}
\caption{Binary star composed of component A being the massive body,
and component B considered to be the light-source.}
\label{FIG: Distance1}
\end{figure}

A scheme of light propagation of a signal emitted at component B in the gravitational field of component A is shown in 
FIG.~\ref{FIG: Distance1}. The vector $\ve{x}_1$ points from the mass center of massive body to the observer, and vector 
$\ve{x}_0$ points from the mass center of massive body to the source, and we define $\ve{R} = \ve{x}_1 - \ve{x}_0$, the 
absolute value $R = \left|\ve{R}\right|$ and unit vector by $\ve{k} = \ve{R}/R$.
Furthermore, the impact vector $\ve{d}=\ve{k} \times \left( \ve{x}_1 \times \ve{k} \right)$ is defined, and its absolute 
value is denoted by $d=\left|\ve{d}\right|$. The Schwarzschild radius of massive body, i.e. of component A of the binary 
system is denoted by $\displaystyle m = \frac{G\,M}{c^2}$.

For determining the light deflection in weak gravitational fields there are two essential approaches: standard 
post-Newtonian approach, e.g. {\it Brumberg} \cite{Brumberg1991}, and classical lens equation 
e.g. Eq.(67) in {\it Fritelli, et al.} \cite{Fritelli_Kling_Newmann}, Eq.~(24) in {\it Bozza} \cite{Bozza}, 
or Eq.~(23) in {\it Zschocke} \cite{Article_Generalized_Lens_Equation}. 
While the first approach is restricted by the condition $m \gg d$, 
the second approach is only valid for the case that source and observer are far from the massive body, especially 
for $a = \ve{k}\cdot \ve{x}_1 \gg d$ and $b = - \ve{k}\cdot \ve{x}_0 \gg d$ (for geometrical illustration of $a$, $b$ 
and $d$ see FIG.~\ref{FIG: Distance1}). However, in binary systems 
extreme configurations are possible like $d = 0$ or $b = 0$. Therefore, in order to investigate the light deflection 
in binary systems one needs a generalized lens equation which is valid in such extreme configurations where 
the standard post-Newtonian approach as well as the classical lens equation cannot be applied. 
Recently, {\it Zschocke} \cite{Article_Generalized_Lens_Equation} has derived a generalized lens equation, which allows to 
determine the light deflection in such extreme astrometric configurations like the binary systems are:
\begin{eqnarray}
\varphi &=& \frac{1}{2}\left(\sqrt{\frac{d^2}{x_1^2} + 8\,\frac{m}{x_1}\,
\frac{x_0\,x_1 - \ve{x}_0\cdot\ve{x}_1}{R\,x_1}} - \frac{d}{x_1}\right).
\label{generalized_lens_1}
\end{eqnarray}

\noindent
Actually, the lens equation has two solutions, but here only one solution is considered, while the second solution 
represents just the second image of one and the same source which is not relevant in our investigation. 
The generalized lens equation is valid up to terms of the order 
$\displaystyle {\cal O}\left(\frac{m^2}{{d^{\;\prime}}^2}\right)$, 
and the absolute value of their total sum can be shown to be smaller or equal to 
$\displaystyle \frac{15\,\pi}{4}\,\frac{m^2}{{d^{\;\prime}}^2}$. 
Here, $\displaystyle d^{\;\prime} = \frac{L}{E}$ is Chandrasekhar's impact parameter, see {\it Chandrasekhar} 
\cite{Chandrasekhar}, where $L$ being the orbital momentum and $E$ is the energy of the photon in the gravitational 
field of massive body. Basically, the light-ray of component B cannot be observed if $d^{\;\prime}$ is smaller than the
radius of massive body A. For stars, the radius is much larger than Schwarzschild radius $m$, hence
$\displaystyle \frac{m^2}{{d^{\;\prime}}^2} \ll 1$. Furthermore, the generalized lens equation (\ref{generalized_lens_1})
is finite for $d \rightarrow 0$ and $b = - \ve{k}\cdot\ve{x}_0 \rightarrow 0$, both of which are possible astrometric 
configurations in binary systems. Furthermore, in {\it Zschocke} \cite{Article_Generalized_Lens_Equation} it has been shown 
that the generalized lens equation (\ref{generalized_lens_1}) yields in the appropriate limits the correct standard 
post-Newtonian solution and the classical lens equation, hence provides a bridge between these essential approaches. 

\begin{figure}[h!]
\centering
\includegraphics[width=8.0cm,angle=270]{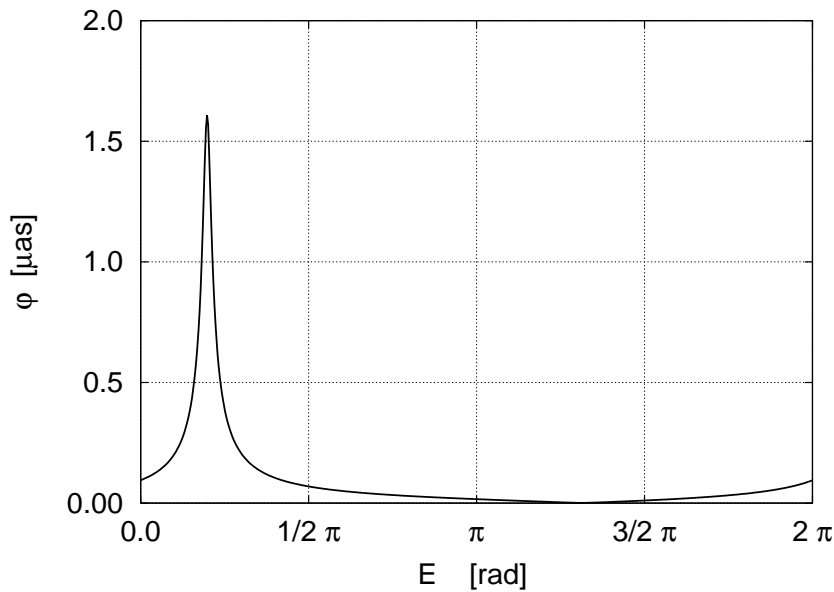}
\caption{Typical light curve of a binary system, determined using generalized lens equation (\ref{generalized_lens_1})
or (\ref{generalized_lens_3}), respectively.
The parameters chosen are: distance $r=1\,{\rm pc}$, semi-major axis $A = 100\,{\rm AU}$
inclination $i=\frac{\displaystyle 31}{\displaystyle 64}\,\pi$, mass $M_A =2\,M_{\odot}$, mass ratio
$\frac{\displaystyle M_A}{\displaystyle M_B}=2.0$, eccentricity ${\rm e}=0.25$, argument of periapsis
$\omega = \frac{\displaystyle \pi}{\displaystyle 4}$.}
\label{FIG: Lightcurve}
\end{figure}

In the following, Eq.~(\ref{generalized_lens_1}) is applied in order to determine the light deflection in binary systems.
For that, the coordinates $\ve{x}_0$ and $\ve{x}_1$ are used in the form as given by Eqs.~(\ref{inclination_10}) and
(\ref{inclination_15}), respectively.
A typical light-curve of a binary system, calculated by means of generalized lens equation (\ref{generalized_lens_1}) is
shown in FIG.~\ref{FIG: Lightcurve}.

Now, an inclination formula is derived from the generalized lens equation (\ref{generalized_lens_1}).
The impact of eccentricity is neglected, circular orbits ${\rm e}=0$ are considered, implying $\omega=0$.
Thus, the coordinates $\ve{x}_0$ and $\ve{x}_1$ are simplified to the expressions (\ref{appendixC_5}) and 
(\ref{appendixC_10}) given in Appendix \ref{AppendixC}.
Furthermore, the maximal value of light deflection of a binary system is of interest, i.e. the astrometric configuration 
$E=0$ is considered here. Then, by inserting these coordinates in the generalized lens equation (\ref{generalized_lens_1}) 
one obtains (see Eq.~(\ref{appendixC_24_C}) in Appendix \ref{AppendixC}) up to terms of the order 
$\displaystyle {\cal O} \left(\frac{A}{r}\,\sqrt{\frac{m}{r}\,\frac{A}{r}}\right)$:
\begin{eqnarray}
\varphi &=& \frac{1}{2}
\left(\sqrt{\frac{A^2}{r^2} \,\cos^2 i  + 8\,\frac{m}{r}\,
\frac{A}{r}\,\left(1 + \sin i\right)} - \frac{A}{r}\,\left|\,\cos i\,\right|\right).
\label{inclination_22}
\end{eqnarray}

\noindent
The minimal value $\varphi_{\rm min}=\varphi \left(i = 0\right)$ and maximal value 
$\varphi_{\rm max}=\varphi \left(i = \frac{\pi}{2}\right)$ of light deflection
for the astrometric position $E=0$ follow from Eq.~(\ref{inclination_22}):
\begin{eqnarray}
\varphi_{\rm min} &=& \frac{1}{2}\left(\sqrt{\frac{A^2}{r^2} + 8\,\frac{m}{r}\,\frac{A}{r}} - \frac{A}{r}\right)
\approx 2\,\frac{m}{r} = 0.0197\,\mu{\rm as}\;\frac{M}{M_{\odot}}\;\frac{\rm pc}{r}\;,
\label{inclination_24}
\\
\nonumber\\
\varphi_{\rm max} &=& 2\,\frac{\sqrt{m\,A}}{r} = 200\,\mu{\rm as}\;
\sqrt{\frac{M}{M_{\odot}}\;\frac{A}{\rm AU}}\;\frac{\rm pc}{r}\;,
\label{inclination_25}
\end{eqnarray}

\noindent
where in (\ref{inclination_24}) terms of the order $\displaystyle {\cal O} \left(\frac{m^2}{r\,A}\right)$ have been 
neglected. The expression (\ref{inclination_22}) can be reconverted in terms of inclination (see Appendix \ref{AppendixD}):
\begin{eqnarray}
\left|\,\frac{\pi}{2} - i\,\right| &=& \arccos \left(- \frac{p}{2} + \sqrt{\frac{p^2}{4} - q } \right),
\label{inclination_30_A}
\end{eqnarray}

\noindent
where
\begin{eqnarray}
p &=& \frac{8\,m^2\,A - 4\,m\,r^2\,\varphi^2}{A \left(r^2\,\varphi^2 + 4\,m^2\right)} \,,
\label{inclination_35}
\\
\nonumber\\
q &=& - \,\frac{A^2\,r^2\,\varphi^2 + 4\,m\,A\,r^2\,\varphi^2-4\,m^2\,A^2 - r^4\,\varphi^4}
{A^2 \left(r^2\,\varphi^2 + 4\,m^2\right)}\,.
\label{inclination_40}
\end{eqnarray}

\noindent
The inclination formula (\ref{inclination_30_A}) yields the upper limit for 
$\displaystyle \left|\,\frac{\pi}{2} - i\,\right|$ of a binary system in order to have a given value of light deflection 
$\varphi$. Note, that the values of $\varphi$ cannot be chosen arbitrarily, but they are restricted by 
$\varphi_{\rm min}$ and $\varphi_{\rm max}$ given by Eqs.~(\ref{inclination_24}) and (\ref{inclination_25}), respectively.

The inclination formula (\ref{inclination_30_A}) can considerably be simplified.
From (\ref{inclination_35}) and (\ref{inclination_40}) one obtains by series expansion
\begin{eqnarray}
p &=& - 4\,\frac{m}{A} + 8\,\frac{m^2}{r^2\,\varphi^2} + {\cal O} \left(m^3\right),
\label{comparison_10}
\\
\nonumber\\
q &=& - 1 - 4\,\frac{m}{A} + 8\,\frac{m^2}{r^2\,\varphi^2} - 4\,\frac{m^2}{A^2} + \frac{r^2\varphi^2}{A^2}
+ {\cal O} \left(m^3\right).
\label{comparison_15}
\end{eqnarray}

\noindent
By means of (\ref{inclination_25}), the last term in (\ref{comparison_15}) can be estimated to be smaller than
$\displaystyle 4\,\frac{m}{A} \ll 1$. Here, it should be underlined that
$\frac{\displaystyle m}{\displaystyle A} \ll \frac{\displaystyle m}{\displaystyle r\,\varphi}$
even at large distances $r \simeq 10^3 \,{\rm pc}$ and small values for semi-major axis $A \simeq 1\,{\rm AU}$.
Thus, one obtains 
\begin{eqnarray}
\left| \frac{\pi}{2} - i \right| &\approx&  \arccos \left( 1 - 8\, \frac{m^2}{r^2\,\varphi^2} \right) 
\approx 2\,\arctan \left(2\,\frac{m}{r\,\varphi}\right)
\label{inclination_49}
\end{eqnarray}

\noindent
up to terms of the order $\displaystyle {\cal O} \left(\frac{m^3}{r^3\,\varphi^3}\right)$ and 
$\displaystyle {\cal O} \left(\frac{m}{A} \right)$; here 
the relation $\arccos \left(1-8\,x^2\right) = 2\,\arctan 2\,x + {\cal O} \left(x^3\right)$ for $x \ll 1$ has been used. It 
should be underlined, that the applicability of (\ref{inclination_49}) is restricted by the condition (\ref{Appendix_KMS_E}) 
given in Appendix \ref{Appendix_KMS} and by $d \gg m$. Here, 
it should be noticed that $x = 2\,\frac{\displaystyle m}{\displaystyle r\,\varphi} \ll 1$, even in such an extreme case 
like $r=1\,{\rm pc}$, $m=m_{\odot}$ and $\varphi \simeq 1\,\mu{\rm as}$ one obtains a small number $x = 0.019$;
for an analytical proof use the exact expression for $\varphi_{\rm min}$. Due to
$\frac{\displaystyle m}{\displaystyle A} \ll \frac{\displaystyle m}{\displaystyle r\,\varphi}$, the impact of
semi-major axis is of lower order and can be neglected in the inclination formula. 
The inclination formula (\ref{inclination_49}) can be expressed in terms of dimensionless quantities as follows:
\begin{eqnarray}
\left| \frac{\pi}{2} - i \right| &\approx& 2\,\arctan \left(0.0197\;\frac{M}{M_{\odot}}\;\frac{\mu{\rm as}}{\varphi}\;
\frac{\rm pc}{r}\right).
\label{inclination_50}
\end{eqnarray}

\noindent
Note, that expression (\ref{inclination_50}) agrees with an inclination formula (\ref{KMS_D}) derived at the first time 
by {\it Klioner, et al.} \cite{inclination_formula}; the arguments of their work are represented 
in Appendix \ref{Appendix_KMS}. The simplified inclination formula (\ref{inclination_50}) is not only useful for
straightforward estimations about the order of magnitude, but it's simple structure provides also an obvious 
comprehension about the interplay of the individual terms.

\section{Stringent conditions on orbital parameters for binary systems\label{Condition1}}

In this Section, two stringent conditions on the orbital elements are highlighted for binaries having a given 
light deflection. These strict conditions are valid for any binary system: astrometric, spectroscopic, 
eclipsing and resolved binaries. 

The first stringent condition follows from the maximal light deflection angle (\ref{inclination_25}), given by
\begin{eqnarray}
\varphi &\le& 200\,\mu{\rm as}\;\sqrt{\frac{M}{M_{\odot}}\;\frac{A}{\rm AU}}\;\frac{\rm pc}{r}\,.
\label{condition_1}
\end{eqnarray}

\noindent
It represents a strict criterion for the maximal light deflection of a binary system with given
Schwarzschild radius of component A, given semi-major axis $A$, and given distance $r$ between
binary system and the observer.

The second stringent condition on the orbital elements follows from the inclination formula in the
simplified form as given by Eq.~(\ref{inclination_50}). For a better illustration, this
condition is given in terms of angular degrees instead of radians. Using $\arctan x = x + {\cal O} (x^3)$ one obtains 
\begin{eqnarray}
\left| \,90^{\circ} - i\,\right| &\le& 2.25^{\circ}\;\frac{M}{M_{\odot}}\;\frac{\mu{ \rm as}}{\varphi}\;\frac{\rm pc}{r}\;.
\label{condition_2}
\end{eqnarray}

\noindent
According to this strict condition, the inclination $i$ of a binary system with mass $M$ (recall $M$ is the stellar mass 
of component A) and at distance $r$ must not exceed the given value in order to have a light deflection $\varphi$.

Both these stringent conditions Eqs.~(\ref{condition_1}) and (\ref{condition_2}) were already stated in the introductory 
Section by Eqs.~(\ref{classical_lens_2}) and (\ref{condition_0}), respectively. However, is should be underlined here, that 
both (\ref{condition_1}) and (\ref{condition_2}) were obtained with the aid of one and the same approach, namely the 
generalized lens equation, while (\ref{classical_lens_2}) and (\ref{condition_0}) were obtained by means of the classical 
lens equation and the post-Newtonian solution, that means by two different approaches. 

The observability of light deflection effect in binaries implies the realization of both these stringent conditions
(\ref{condition_1}) and (\ref{condition_2}) simultaneously for a given binary system. But even if a given binary
system satisfies these both conditions, the observability of light deflection effect is not guaranteed, because
the astrometric position $E=0$ has to be reached during the time of observation. Nonetheless, as soon as the
orbital elements $r$, $A$, $m$ and $i$ of the binaries are known, these both stringent conditions (\ref{condition_1})
and (\ref{condition_2}) allow to find a possible candidate for being a relevant binary system for a given 
light deflection $\varphi$ depending on the instrumentation of the observer. However, as it will be shown in 
Section \ref{Section3a} and Section \ref{Section3b}, the existence of such systems is highly improbable.


\section{Total number of binaries with a given light deflection for infinite time of observation}\label{Section3a}

In the previous Section, the conditions on orbital parameters for a binary system have been determined in order to
have a given magnitude of light deflection $\varphi$. In this Section, the total number of such relevant 
binaries is determined. In order to estimate the total number of binaries having a given light deflection $\varphi$, the 
following formula is applied:
\begin{eqnarray}
N \left(\varphi\right) &=& \int\limits_{R_{\rm min}}^{R_{\rm max}} d^3 r\;\rho (r)\;
\int\limits_{A_{\rm min}}^{A_{\rm max}} d A\; f (A)\;
\int\limits_{\mu_{\rm min}}^{\mu_{\rm max}} d \mu \; f (\mu)\;P(i)\,.
\label{numerical_results_5}
\end{eqnarray}

\noindent
Here, $\rho (r)$ is the density of binaries, $f(A)$ is the semi-major axis distribution of binary systems,
$f(\mu)$ is the mass distribution of stars where $\displaystyle \mu = \frac{M}{M_{\odot}}$ is the mass-ratio of
massive body (component A) and solar mass. The probability distribution $P(i)$ to find a binary system with given
inclination $0 \le i \le \pi$, is a function of distance $r$, semi-major axis $A$, mass ratio $\mu$, and the given light 
deflection angle $\varphi$. According to (\ref{inclination_30_A}), the probability distribution $P(i)$ is given by
(the inclination of binary systems is of course a random distribution)
\begin{eqnarray}
P(i) &=& \frac{2}{\pi}\,\arccos \left(- \frac{p}{2} + \sqrt{\frac{p^2}{4} - q } \right),
\label{inclination_30_B}
\end{eqnarray}

\noindent
where $p$ and $q$ are given by Eqs.~(\ref{inclination_35}) and (\ref{inclination_40}).

For the minimal distance of a binary system from the Sun one can safely assume $R_{\rm min} = 1\,{\rm pc}$.
From (\ref{inclination_25}) follows that beyond a sphere with radius $R_{\rm max} = 2000\,{\rm pc}$ only a very few
exceptional binary systems might have a light deflection of more than $1\,\mu{\rm as}$. The Sun is located in the
{\it Orion arm} which has about $1000\,{\rm pc}$ across and approximately $3000\,{\rm pc}$ in length.
For the estimate according to (\ref{numerical_results_5}), it is meaningful to assume that the stars are homogeneously 
distributed inside the {\it Orion arm}. For the uniform star density $\rho_{\rm stars} = 0.1\,{\rm star}\;{\rm pc}^{-3}$,
a value which is in line with the data of {\it RECONS} \cite{Recons}.
Furthermore, a common presumption is, that about $50$ percent of all stars are
components of a binary or multiple system, see {\it Duquennoy {\rm \&} Mayor} \cite{Binary1} 
and {\it Halbwachs, et al.} \cite{Binary2}. Then one obtains for the density of binaries
\begin{eqnarray}
\rho (r) &\simeq& 0.025\; {\rm binaries}\;{\rm pc}^{-3}\,.
\label{density_10}
\end{eqnarray}

\noindent
Let us now consider the distribution of semi-major axis $A$ in binary systems. Statistical investigations show that the
distribution of binary semi-major axis is flat in a logarithmic scale over the range of six orders of magnitude, that means
assumed to be valid in the large range $A_{\rm min} = 10\,R_{\odot} \le A \le  10^4\,{\rm AU} = A_{\rm max}$,
see {\it Kouwenhoven {\rm \&} de Grijs} \cite{orbital_distribution6}. 
The lower limit ${A}_{\rm min}$ is determined by the semi-major axis at which Roche lobe
overflow occurs, while the upper limit ${\rm A}_{\rm max}$ depends on how large the averaged star density is. The
logarithmic distribution is known as "{\it \"Opik's law}" (1924) named after its discoverer, and is given by
$\displaystyle f(A) \sim \frac{1}{A}$, a law which has also been confirmed by recent investigations, 
see {\it Poveda, et al.} \cite{semi_major_axis_5}. This distribution
is a consequence of the process of star formation as well as of the dynamical history of the binary system, and one can 
take this law as a given fact for numerical studies. Accordingly, see Appendix \ref{Probability_Distribution}: 
\begin{eqnarray}
f (A) &=& \frac{1}{A}\;
\left(\ln \frac{A_{\rm max}}{A_{\rm min}}\right)^{-1}\,.
\label{major_axis_7}
\end{eqnarray}

\noindent
Furthermore, for the mass distribution $f \left(\mu\right)$ let us recall the initial mass function (IMF) which is the
probablity that a star is newly formed with a stellar mass $M$ and is frequently assumed to be a power law
$f (M) \sim M^{-\alpha}$.  Originally, the IMF has been introduced by {\it Salpeter} \cite{IMF1} for solar
neighborhood region, who assumed the value $\alpha = 2.35$ and a validity region for stars with masses between
$0.4\,M_{\odot}$ and $10\,M_{\odot}$. During the past decades the IMF has been refined by
several investigations. Especially, the numerical values of slope parameter $\alpha$ and regions of validity have been
proposed in subsequent investigations, e.g. {\it Scalo} \cite{IMF2}, {\it Robin, et al.} \cite{IMF3}. 
{\it Ninkovic {\rm \&} Trajkovska} \cite{massdistribution1}, {\it Ninkovic} \cite{massdistribution2}, 
{\it Kroupa} \cite{massdistribution3}; for a review see {\it Kroupa} \cite{review_IMF}. 
Moreover, the IMF does not necessarily coincide with the real mass distribution 
of stars, because IMF describes mass distribution of a star formation, while the
solar neighborhood mainly consists of evolved stars of main sequence. Here, for simplicity this distribution is used 
as a given fact with $\alpha=2.35$, and the proposed region of validity $\mu_{\rm min} = 0.4$, and $\mu_{\rm max} = 10$ 
is assumed. According to IMF, one finds for $\alpha \neq 1$
(see Appendix \ref{Probability_Distribution})
\begin{eqnarray}
f (\mu) &=&
\frac{(1 - \alpha)\;\mu^{-\alpha}}{\mu_{\rm max}^{(1-\alpha)} - \mu_{\rm min}^{(1-\alpha)}} \,.
\label{mass_distribution_7}
\end{eqnarray}

\begin{figure}[h!]
\centering
\includegraphics[width=8.0cm,angle=270]{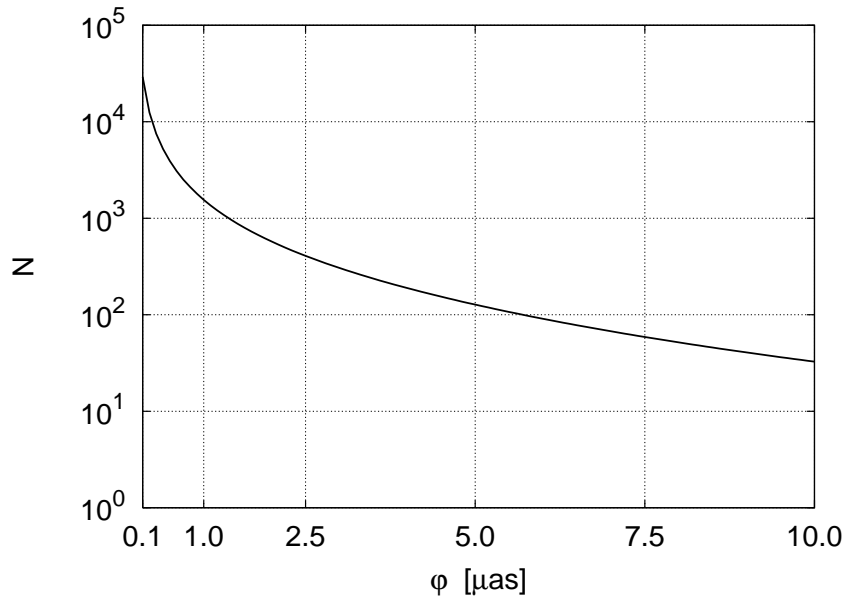}
\caption{Total number of binaries according to Eq.~(\ref{numerical_results_5}), having parameters such that the
light deflection of component B at component A is larger than a given value for $\varphi$. Note, that 
for one individual observation the astrometric accuracy of Gaia is about $25\,\mu{\rm as}$ and the end-of-mission 
accuracy is about $5\,\mu{\rm as}$ in the most ideal case (bright star). The observation time is assumed to be infinite, 
so that it is guaranteed that the optimal configuration $E=0$ is reached.}
\label{FIG: Number_Binaries_2}
\end{figure}

\noindent
In order to motivate that distribution further, one can compare (\ref{mass_distribution_7}) with the {\it RECONS} data
\cite{Recons} where one finds a fair agreement. Using (\ref{inclination_30_B}) - (\ref{mass_distribution_7}), the results
of the estimate (\ref{numerical_results_5}) are shown in FIG.~\ref{FIG: Number_Binaries_2}; recall, when evaluating this 
integral one has to take into account the boundary conditions given by Eqs.~(\ref{inclination_24}) and 
(\ref{inclination_25}).

 According to
FIG.~\ref{FIG: Number_Binaries_2}, in total there are about $N \sim 10^3$ binaries having a light deflection of at least
$\varphi = 1\,\mu{\rm as}$. Here, the observation time is assumed to be infinite,
so that it is guaranteed that the optimal configuration $E=0$ is reached.

\section{Total number of binaries with a given light deflection for finite time of observation}\label{Section3b}

In Eq.~(\ref{numerical_results_5}) the number of binaries with a given maximal possible light deflection $\varphi$ has been 
determined, just by taking for eccentric anomaly the value $E=0$, that means the ideal
configuration where the light deflection takes its maximal value (note, the eccentricity $e=0$). It is, however, obvious that
during the most part of the orbital motion one will have $E \neq 0$ and the light deflection will be much smaller than
the maximal possible light deflection angle $\varphi$. On the other side, the orbital period $T$ of relevant binaries, 
given by Eq.~(\ref{appendixB_55}), will easily exceed the time of observation; for instance, the Gaia mission 
time is about $T_{\rm mission} \simeq 5\,{\rm years}$. Therefore, it will be not very probable,
that the component B will be just at the relevant position near the value $E=0$, where the light deflection becomes
observable on microarcsecond level. In order to determine that number of observable relevant binaries, one has to extend
Eq.~(\ref{numerical_results_5}) as follows,
\begin{eqnarray}
N \left(\varphi\right) &=& \int\limits_{R_{\rm min}}^{R_{\rm max}} d^3 r \; \rho (r)
\int\limits_{A_{\rm min}}^{A_{\rm max}} d A \;f(A)
\int\limits_{\mu_{\rm min}}^{\mu_{\rm max}} d \mu \; f(\mu)\;P(i)\;P(E).
\label{observable_5}
\end{eqnarray}

\noindent
Here, $P(E)$ is the probability for the binary system to be in the region $E$, where the light
deflection is larger than a given value for $\varphi$.

\begin{figure}[h!]
\centering
\includegraphics[width=8.0cm,angle=270]{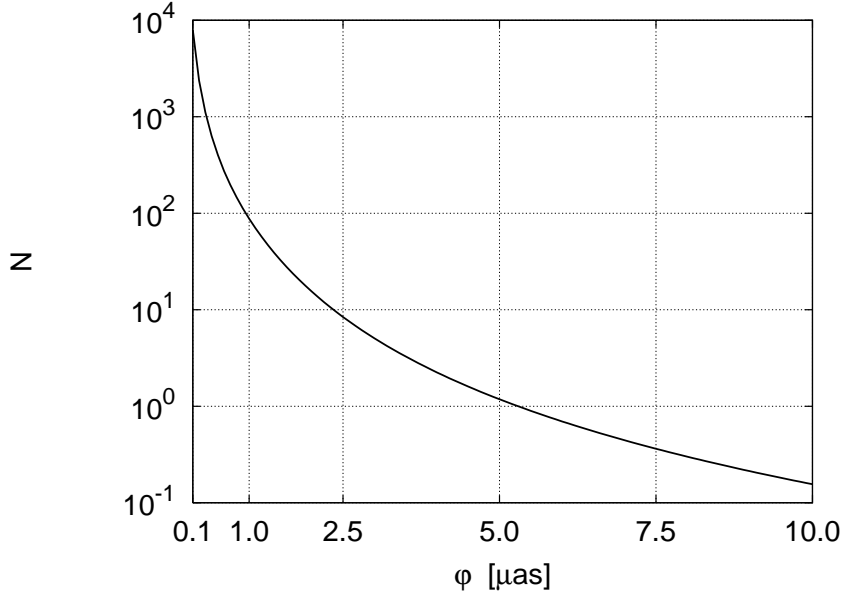}
\caption{Total number of binaries according to Eq.~(\ref{observable_5}) where the binary system
reaches the optimal configuration $E=0$ during an assumed observation time of $T_{\rm mission} = 5\,{\rm years}$, 
and having orbital parameters such that the light deflection of component B at component A
is larger than a given value for $\varphi$. Note, that for one individual observation the astrometric accuracy of Gaia is 
about $25\,\mu{\rm as}$ and the end-of-mission accuracy is about $5\,\mu{\rm as}$ in the most ideal case (bright star).}
\label{FIG: Number_Binaries_3}
\end{figure}

In the very same way, as applied for the derivation of the inclination formula (\ref{inclination_30_A}), one can 
reconvert (\ref{appendixC_24_D}) in terms of eccentric anomaly $E$ and one finds the eccentric anomaly formula:
\begin{eqnarray}
E &=& \pm \; \arccos \left( - \frac{p}{2} + \sqrt{\frac{p^2}{4} - q }\right),
\label{observable_30}
\end{eqnarray}

\noindent
where $p$ and $q$ are given by Eqs.~(\ref{inclination_35}) and (\ref{inclination_40}).
For a given value of light deflection $\varphi$, the formula (\ref{observable_30}) yields the value
of eccentric anomaly $E$ of a binary system characterized by semi-major axis $A$ and Schwarzschild radius (or mass) $m$ at 
a distance $r$. However, the values of $\varphi$ cannot be chosen arbitrarily, instead they are restricted by
$\displaystyle \varphi_{\rm min}=\varphi \left(E=\pm \frac{\pi}{2}\right)$ and 
$\varphi_{\rm max}=\varphi \left(E=0\right)$ given by (of course, only astrometric positions with
$\displaystyle 0 \le E \le \frac{\pi}{2}$ are taken into account, because for the area 
$\displaystyle \frac{\pi}{2} \le E \le \pi$ the light deflection is negligible):
\begin{eqnarray}
\varphi_{\rm min} &=& \frac{1}{2}\left(\sqrt{\frac{A^2}{r^2} + 8\,\frac{m}{r}\,\frac{A}{r}} - \frac{A}{r}\right)
\approx 2\,\frac{m}{r} = 0.0197\,\mu{\rm as}\;\frac{M}{M_{\odot}}\;\frac{\rm pc}{r}\;,
\label{observable_20}
\\
\varphi_{\rm max} &=& 2\,\frac{\sqrt{m\,A}}{r} = 200\,\mu{\rm as}\;
\sqrt{\frac{M}{M_{\odot}}\;\frac{A}{\rm AU}}\;\frac{\rm pc}{r}\;,
\label{observable_25}
\end{eqnarray}

\noindent
where in (\ref{observable_20}) terms of the order $\displaystyle {\cal O} \left(\frac{m^2}{r\,A}\right)$ have been 
neglected. These expressions resemble the corresponding expressions in 
Eqs.~(\ref{inclination_24}) and (\ref{inclination_25}).
According to Eq.~(\ref{observable_30}), the region where the binary system has a light deflection larger or equal 
$\varphi$ is given by $2\,E$. One has also to take into account, that during Gaia mission time $T_{\rm mission}$
the component B moves along the orbit and could move into the region $2\,E$. Therefore, the probability $P(E)$
that the binary system is during Gaia mission time at least ones inside the relevant astrometric position with the
value $E$ in (\ref{observable_30}), is given by
\begin{eqnarray}
P\left(E\right) &=& {\cal P}_1 \left( \frac{1}{\pi}\; \arccos \left( - \frac{p}{2} + \sqrt{\frac{p^2}{4} - q }\right)
+ \frac{T_{\rm mission}}{T} \right)\,,
\label{observable_35}
\end{eqnarray}

\noindent
where the operator ${\cal P}_1$ is defined by 
\begin{eqnarray}
{\cal P}_1 \left( x\right) =
\begin{array}[c]{c}
\displaystyle
x \quad {\rm if} \quad x < 1\,,\\
\displaystyle
1 \quad {\rm if} \quad x \ge 1\,.
\end{array}
\label{observable_37}
\end{eqnarray}

\noindent
The probability distribution (\ref{observable_35}) has to be implemented in Eq.~(\ref{observable_5})
in order to determine the number of binary systems having a given light deflection $\varphi$ and
to be observable during Gaia mission time $T_{\rm mission}$. The results of Eq.~(\ref{observable_5}) are shown in
FIG.~\ref{FIG: Number_Binaries_3}; recall, when evaluating this integral one has to take into account the boundary 
conditions given by Eqs.~(\ref{inclination_24}) and (\ref{inclination_25}). Accordingly, there are only a very few binaries
$\sim 10^2$ having a light deflection of at least $\varphi = 1\,\mu{\rm as}$, if the astrometric position $E=0$ 
is reached during an assumed observation time of $T_{\rm mission} = 5\,{\rm years}$.


\section{Special case: conditions on orbital parameters for resolved binaries\label{Condition2}}

In this Section, the special case of a resolved binary system is considered. In order to investigate the detectability of 
the light deflection effect in binaries, todays most modern astrometric mission, the ESA cornerstone mission Gaia, 
and its instrumentation which provides the highest possible astrometric accuracy at the moment, will be considered here 
as a concrete example. 

\subsection{Resolving power of Gaia}\label{Section4}

\begin{figure}[h!]
\centering
\includegraphics[width=12.0cm,angle=270]{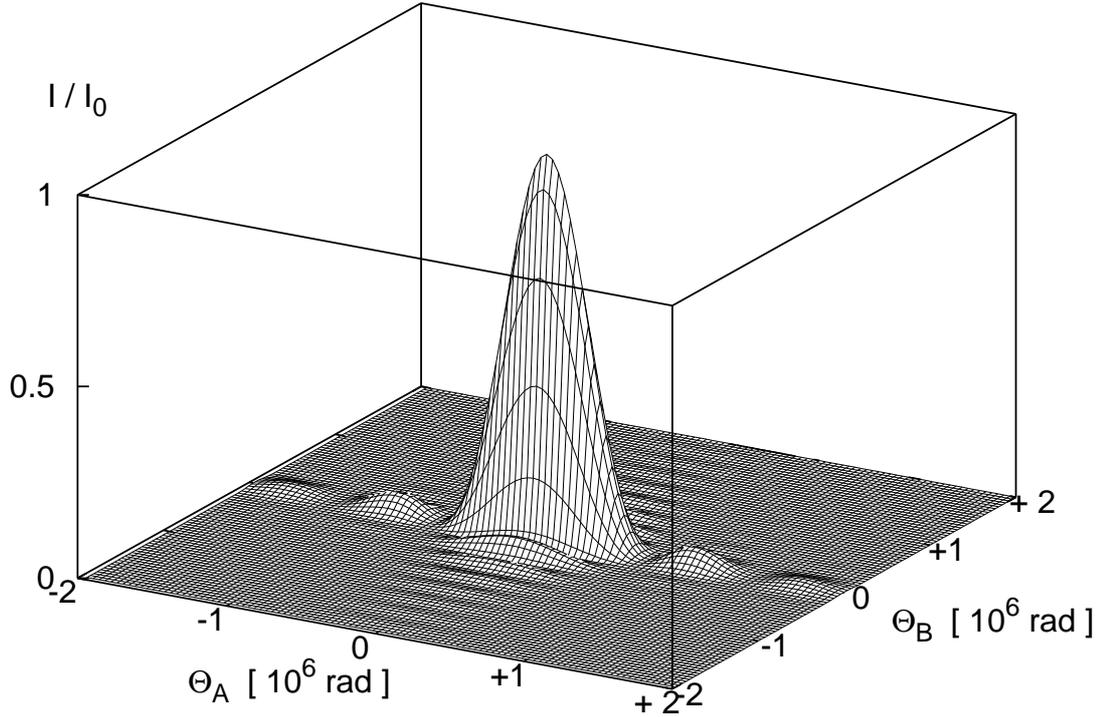}
\caption{Point Spread Function (PSF) for a rectangular telescope according to Eq.~(\ref{rectangular_5}).
The incident monochromatic light-ray has a wavelength 
of $\lambda = 350\,{\rm nm}$. The parameters of the rectangular telescope are: ${\rm A} = 0.5\,{\rm m}$,
${\rm B} = 1.45\,{\rm m}$.}
\label{FIG: Resolving_Power1}
\end{figure}

The core of Gaia optical instrumentation consists of two identical mirror telescopes, ASTRO-1 and ASTRO-2, with a
rectangular pupil whose dimensions are ${\rm A}=0.50\,{\rm m}$, ${\rm B}=1.45\,{\rm m}$, and $f=35\,{\rm m}$ is the
effective focal length. The intensity is given by, see {\it Lattanzi, et al.} \cite{PSF1} and {\it Lindgren} \cite{PSF2}:
\begin{eqnarray}
I(z_{\rm A},z_{\rm B}) &=& I_0 \left( \frac{{\rm sin}^2 (z_{\rm A})}{z_{\rm A}^2}\;
\frac{{\rm sin}^2 (z_{\rm B})}{z_{\rm B}^2} \right)\,,
\label{rectangular_5}
\end{eqnarray}

\noindent
where $z_{\rm A} = \pi\;{\rm A} / \lambda\,{\rm sin} \Theta_{\rm A}$,
$z_{\rm B} = \pi\;{\rm B} / \lambda\,{\rm sin} \Theta_{\rm B}$, ${\rm A}$ and ${\rm B}$ the width and length of rectangular
mirror, $\lambda$ is the wavelength of incident light-ray, and $\Theta_{\rm A}, \Theta_{\rm B}$ are the angle of
observation, i.e. the angle between the axis of the rectangular aperture and the line between aperture center and
observation point. The intensity of incident light-ray at $\Theta_{\rm A} = 0, \Theta_{\rm B}=0$ is denoted by $I_0$.
The function $I(z_{\rm A},z_{\rm B})/I_0$ in Eq.~(\ref{rectangular_5}) is the (by $\Theta_{\rm A}=\Theta_{\rm B}=0$
normalized) Point Spread Function (PSF) for monochromatic incident light with wavelength $\lambda$ for a rectangular
aperture. The optical spectrum of stars is $\lambda = \left(350 - 750\right) \, {\rm nm}$.
In FIG.~\ref{FIG: Resolving_Power1} the PSF for an incident monochromatic light-ray with $\lambda=350\,{\rm nm}$
is represented for Gaia telescopes.

\noindent
Most of the light is concentrated in the central bright rectangular shaped pattern. The length $l_{\rm A}$ and width
$l_{\rm B}$ of this rectangle is determined by the first zero-roots of (\ref{rectangular_5}) at $z_{\rm A} \simeq \pi$ and
$z_{\rm B} \simeq \pi$, respectively. From that it follows, that ${\rm sin} \Theta_{\rm A} = \pi \;\lambda /(\pi\;{\rm A})
= \lambda / {\rm A}$, and ${\rm sin} \Theta_{\rm B} = \pi \;\lambda /(\pi\;{\rm B}) = \lambda / {\rm B}$. Furthermore, if
the diffraction pattern is shown on a screen at a distance $f$, then the length and width is given by, see 
{\it Hog, et al.} \cite{optical_design}: 
\begin{eqnarray}
L_{\rm A} &=& 2\;\frac{f\;\lambda}{\rm A}\,,
\label{rectangula_10}
\\
\nonumber\\
L_{\rm B} &=& 2\;\frac{f\;\lambda}{\rm B}\,,
\label{rectangula_15}
\end{eqnarray}

\noindent
where $f$ is the focal length of the optic, i.e. of the rectangular Gaia mirror.
Accordingly, the given numerical values ${\rm A}=0.50\,{\rm m}$, ${\rm B}=1.45\,{\rm m}$,
and $\lambda = 350 \;{\rm n} {\rm m}$ result in $L_{\rm A} = 49.0\;\mu{\rm m}$ and $L_{\rm B} = 16.9\;\mu{\rm m}$
for the length and width of "{\it Airy rectangle}" of Gaia optics. Note, that the
"{\it Airy rectangle}" has the same order of magnitude than the pixel size ($10 \mu{\rm m} \times 30 \mu{\rm m}$) of the 
$110$ CCDs (Charge-Coupled Device) sensors of astrometric field part of the focal plane. In order to separate two pointlike 
sources, the distance between their centers of the rectangle has to be larger than either $L_{\rm A}$ or $L_{\rm B}$.
Since $L_{\rm B} < L_{\rm A}$, in this study the better resolution value $L_{\rm B}$ is used,
which corresponds to a resolution angle of
\begin{eqnarray}
\delta &=& \frac{L_{\rm B}}{2\,f} = \frac{\lambda}{\rm B} \,.
\label{rectangula_30}
\end{eqnarray}

\noindent
The resolving power is the minimal angular distance between two objects to get separable by Gaia instrumentation.
With the parameters given above one obtains the resolving power $\delta$ of Gaia optics:
\begin{eqnarray}
\delta &=& 0.24 \times 10^{-6}\;{\rm rad} = 49.7 \; {\rm mas} \,.
\label{minimal_alpha_10}
\end{eqnarray}

\noindent
In what follows this parameter is of fundamental importance in order to determine the ability
of Gaia to determine the light deflection in binary systems.

\subsection{Orbital Parameters of Resolved Binaries Observable by Gaia}\label{Section5}

In this Section, the question is addressed which and how many binary systems can be separated by Gaia instrumentation
among all those relevant binaries found in the previous Section; see FIG.~\ref{FIG: Number_Binaries_3}.
In average, Gaia will observe each object $80$ times, but will not constantly observe these objects during mission time.
However, for simplicity the scanning law of Gaia is approximated by assuming a permanent observation of all objects
during the whole mission time.

Furthermore, visual binaries are considered in this Section, i.e. binaries which are separable by Gaia telescopes.
The both largest telescopes of Gaia have a resolution angle $\delta$ discussed in the previous Section,
see Eqs.~(\ref{rectangula_30}) and (\ref{minimal_alpha_10}).
For binary systems, this resolution angle $\delta$ corresponds to a minimal distance between the components A and B
to get separable within Gaia optics. Using (\ref{expression_d}) and (\ref{rectangula_30}) one obtains the condition
\begin{eqnarray}
d &=& A\,\left| \cos i \right| \,\ge\, \delta \, r \,=\, \frac{\lambda}{B} \, r \,,
\label{optics_20}
\end{eqnarray}

\noindent
where $r$ is the distance between the remote objects and Gaia observer. This condition is by far
much more important than taking into account the effect of finite radius of the stars, which would imply
$A\,\left| \cos i \right| \,\ge\, R_A$, where $R_A$ being the radius of component $A$. By inserting the extreme case
$A \,\left| \cos i \right| = \delta\,r$ into (\ref{inclination_22}), one obtains
\begin{eqnarray}
\varphi &=& \frac{1}{2}
\left( \sqrt{\delta^2 + 8\left(1+\sin i\right)\,\frac{m}{r}\,\frac{A}{r}} - \delta\right) 
\simeq 4\,\frac{m}{r}\,\frac{A}{r}\,\frac{1}{\delta} \,, 
\label{visual_5}
\end{eqnarray}

\noindent
where $\sin i \simeq 1$ has been used, and terms of higher order ${\cal O} \left(m^2\right)$ are neglected.
Relation (\ref{visual_5}) is an expression for the maximal light deflection angle of a binary system when taking
into account the resolving power of Gaia. Eq.~(\ref{visual_5}) is a much stricter restriction than the generalized
lens equation (\ref{inclination_22}), because (\ref{visual_5}) determines the light deflection angle only of those
binary systems having a resolution angle $\delta$ of Gaia optical instrumentation, while (\ref{inclination_22})
determines the light deflection angle of any possible binary system. Note, that from Eq.~(\ref{visual_5}) follows
the maximal possible distance of visual binaries for a given light deflection:
\begin{eqnarray}
r &\le& \sqrt{4\,\frac{m\,A}{\varphi\,\delta}} \,=\,0.18\,{\rm pc}\;\sqrt{\frac{M}{M_{\odot}}\,\frac{A}{\rm AU}}\,,
\label{condition_3}
\end{eqnarray}

\noindent
where in the last expression the optimal values $\delta=0.24\times 10^{-6}\,{\rm rad}$
and $\varphi = 25\,\mu{\rm as}$ has been used. The condition (\ref{condition_3}) can also be written by
\begin{eqnarray}
A &\ge& 30\,{\rm AU}\;\frac{M_{\odot}}{M}\;\frac{r^2}{{\rm pc}^2}\;.
\label{condition_4}
\end{eqnarray}

\noindent
These both conditions (\ref{condition_3}) and (\ref{condition_4}) imply rather extreme orbital
parameters on visual binaries. For instance, condition (\ref{condition_3}) implies
a maximal distance of $r \le 18\,{\rm pc}$ for solar-mass type binaries even with a huge semi-major axis of
$A = 10^4\,{\rm AU}$, while condition (\ref{condition_4}) implies a large semi-major axis for solar-mass type binaries 
outside a sphere of $r \ge 10\,{\rm pc}$. It is almost certain, that such extreme parameters will not be realized in nature.

\section{Summary}\label{Summary}

In this study, the light deflection in binary systems has been considered. While there is absolutely not any doubt about 
the existence of this relativistic effect, is has not been observed so far. 
To investigate this effect of light deflection, an inclination formula (\ref{inclination_30_A}) has been derived by means 
of generalized lens equation (\ref{generalized_lens_1}) obtained recently by {\it Zschocke} 
\cite{Article_Generalized_Lens_Equation}, and these both equations are the theoretical basis for investigating 
the light deflection effect in binary systems. A simplified inclination formula has been presented by 
Eq.~(\ref{inclination_50}) and its validity has been discussed in some detail. This simplified inclination formula has 
also been obtained by {\it Klioner, et al.} \cite{inclination_formula} by an independent approach. 
Furthermore, two stringent conditions on the orbital parameters have been given by Eqs.~(\ref{condition_1}) and 
(\ref{condition_2}). These both stringent conditions are relations between the orbital elements 
of a (resolved, astrometric, eclipsing, spectroscopic) binary system for a given magnitude 
of light deflection, and allow to find a relevant binary system in a straightforward way.

In Section \ref{Section3a}, the total number
of binaries with a given light deflection has been determined by means of
the semi-major axis distribution according to "{\it \"Opik's law}" and the
mass distribution according to "{\it Salpeter's mass distribution}". 
Since the inclinations are randomly distributed, the
inclination formula allows to estimate the total number of relevant
binaries with the aid of Eq.~(\ref{numerical_results_5}). It turns out, that in total 
there exist about $10^3$ binaries having orbital parameters such that the light deflection 
amounts to be at least $1\,\mu{\rm as}$; see FIG.~\ref{FIG: Number_Binaries_2}.

In Section \ref{Section3b} a finite time of observation of $5$ years (Gaia mission time) has been considered, which
considerably reduces the total number of relevant binaries. Clearly, this case is of
more practical importance, since a restricted time window of observation is in more
agreement with reality than the first scenario. By taking into account the probability
to find the system in the ideal astrometric position $E=0$ where the light
deflection becomes maximal, it has been found by evaluating the corresponding
integral (\ref{observable_5}) that there is not any of the relevant binary systems 
in the ideal position $E=0$ during Gaia mission time; see FIG.~\ref{FIG: Number_Binaries_3}. 
Thus, while in principle a few binaries will have a significant light deflection, the effect could not be detected 
due to the restricted finite time window of observation. 

Furthermore, the special case of resolved binaries has been considered in Section
\ref{Condition2}. The astrometric instrumentation of the ESA cornerstone
mission Gaia, see e.g. {\it Perryman, et al.} \cite{Gaia_Overview}, has
been considered in some detail in order to decide whether or not this
subtle effect of light deflection can be observed. Two conditions for
resolved binaries were presented in Eqs.~(\ref{condition_3}) and
(\ref{condition_4}) for such special kind of binary systems. It has been
shown, however, that even for the Gaia mission, which is an outstanding
milestone of progress in astrometry, such binary systems must have rather
extreme orbital parameters in order to reach todays level of
detectability, i.e. on microarcsecond level. The existence of such exotic
binaries is, however, highly improbable.

In summary, the main results are presented by the inclination formulae (\ref{inclination_30_A}) and its simplified version 
(\ref{inclination_50}), the stringent conditions (\ref{condition_1}) and (\ref{condition_2}) and by the
diagrams FIG.~\ref{FIG: Number_Binaries_2} and FIG.~\ref{FIG: Number_Binaries_3}. 
Accordingly, one comes to the conclusion that the detectability of light deflection in binary systems reaches the 
technical limit of todays high precision astrometry and might be detected only in case of a very few and highly exotic 
binary systems. It is, however, very unlikely that such extreme binaries might exist. It seems that the detection of the light deflection effect in binary
systems needs an astrometric accuracy of better than about $0.1 \mu{\rm
as}$. Thus, only astrometric missions of the next generation
can accept the challenge to detect this subtle effect of relativity.

\section*{Acknowledgements}
This work was partially supported by the BMWi grants 50\,QG\,0601 and
50\,QG\,0901 awarded by the Deutsche Zentrum f\"ur Luft- und Raumfahrt
e.V. (DLR). Enlighting discussions with Prof. Sergei A. Klioner, 
Prof. Michael H. Soffel and Prof. Francois Mignard are gratefully acknowledged.

\newpage

\begin{appendix}

\section{Two-body problem\label{AppendixB}}

The calculations in this Appendix follow mainly {\it Landau {\rm \&} Lifshitz} \cite{Landau_Lifshitz}.
Consider two massive bodies, one component having a mass $M_A$ and spatial
coordinate $\ve{r}_A$, and second component with a mass $M_B$ and spatial coordinate $\ve{r}_B$, respectively.
They orbit around their common center of mass $\ve{r}_{\rm CMS}$,
\begin{eqnarray}
\ve{r}_{\rm CMS} &=& \frac{1}{M_A + M_B} \left( M_A \;\ve{r}_A + M_B \;\ve{r}_B \right)\,.
\label{appendixB_1}
\end{eqnarray}

\noindent
The Lagrangian ${\cal L}$ of the two-body problem is given by
\begin{eqnarray}
{\cal L} &=& \frac{M_A}{2} \; \dot{\ve{r}}_A^2 + \frac{M_B}{2} \; \dot{\ve{r}}_B^2 - U (|\ve{r}_A - \ve{r}_B|) \,,
\label{appendixB_5}
\end{eqnarray}

\noindent
where $U$ is the potential. With the aid of relative coordinate $\ve{r}_{AB} = \ve{r}_A - \ve{r}_B$ and
reduced mass $\overline{M} = M_A\;M_B / (M_A + M_B)$, the two-body problem can be transformed into
into one-body problem,
\begin{eqnarray}
{\cal L} &=& \frac{\overline{M}}{2} \; \dot{\ve{r}}^2_{AB} - U (r_{AB}) \,.
\label{appendixB_10}
\end{eqnarray}

\noindent
Polar coordinates $(r_{AB},\phi)$ yield 
\begin{eqnarray}
{\cal L} &=& \frac{1}{2} \left( \overline{M} \; \dot{r}^2_{AB} + r^2_{AB}\; \dot{\phi}^2 \right) - U \left(r_{AB}\right).
\label{appendixB_15}
\end{eqnarray}

\noindent
The orbital angular momentum $L$ is conserved
\begin{eqnarray}
L &=& \overline{M} \; r_{AB}^2\; \dot{\phi} = {\rm const}\,,
\label{appendixB_20}
\end{eqnarray}

\noindent
by means of which one obtains for the total energy of the two-body system the expression
\begin{eqnarray}
E &=& \frac{\overline{M}}{2} \;\dot{r}^2_{AB} + \frac{L^2}{2 \;\overline{M} \;r^2_{AB}} + U \left(r_{AB}\right) \,.
\label{appendixB_25}
\end{eqnarray}

\noindent
From Eq.~(\ref{appendixB_25}) one deduces
\begin{eqnarray}
\dot{r}_{AB} &=& \left(\frac{2}{\overline{M}} \left[ E - U \left(r_{AB}\right) \right]
- \frac{L^2}{\overline{M}^2\;r_{AB}^2}\right)\;,
\label{appendixB_30}
\end{eqnarray}

\noindent
and from Eq.~(\ref{appendixB_30}) one obtains up to an integration constant 
\begin{eqnarray}
t &=& \int d r_{AB}\; \left(\frac{2}{\overline{M}} \left[ E - U \left(r_{AB}\right) \right]
- \frac{L^2}{\overline{M}^2\;r^2_{AB}}\right)^{-1/2}\,,
\label{appendixB_35}
\\
\nonumber\\
\phi &=& \int d r_{AB}\; \frac{\overline{M}}{r^2_{AB}}\;
\left( 2 \,\overline{M} \left[ E - U \left(r_{AB}\right) \right]
- \frac{L^2}{r_{AB}^2}\right)^{-1/2}\,,
\label{appendixB_36}
\end{eqnarray}

\noindent
where in the second relation Eq.~(\ref{appendixB_20}) has been used; note that (\ref{appendixB_36}) is the relation 
between $r_{AB}$ and $\phi$ and is called orbital equation. The Eqs.~(\ref{appendixB_35}) and (\ref{appendixB_36}) are 
the general integral solutions of a two-body problem. In order to integrate these Eqs.~(\ref{appendixB_35}) and 
(\ref{appendixB_36}) one has to specify the potential $U$. In case of Kepler problem one has 
\begin{eqnarray}
U(r) &=& - \frac{\alpha}{r_{AB}} \quad {\rm with} \quad \alpha = G\,M_A\,M_B\,.
\label{appendixB_40}
\end{eqnarray}

\noindent
The Eq.~(\ref{appendixB_36}) can be integrated and yields
\begin{eqnarray}
\phi &=& \arccos \left( \frac{L}{r_{AB}} - \frac{\gamma\,\overline{M}\,M_A\,M_B}{L} \right)
\left( 2 \overline{M}\,E + \frac{\gamma^2\,\overline{M}^2\,M_A^2\,M_B^2}{L^2}\right)^{-1/2} \,,
\label{appendixB_41}
\end{eqnarray}

\noindent
where the axes are chosen such that the mentioned integration constant vanishes.
Furthermore, by introducing the eccentricity ${\rm e}$ (possible values of eccentricity are between 
$0 \le {\rm e} < 1$; ${\rm e} = 0$ corresponds to a circular orbit),
\begin{eqnarray}
{\rm e} &=& \left( 1 + \frac{2 \; E \; L^2\;(M_A + M_B)}{\gamma^2\;M_A^3\;M_B^3}\right)^{1/2}\,,
\label{appendixB_45}
\end{eqnarray}

\noindent
the solution (\ref{appendixB_41}) can be written as
\begin{eqnarray}
\frac{1}{r_{AB}}\;\frac{L^2}{\gamma\,\overline{M}\,M_A\,M_B} &=& 1 + {\rm e} \; \cos \phi\,.
\label{appendixB_46}
\end{eqnarray}

\noindent
Note the expressions of semi-major axis $A$ and semi-minor axis $B$,
\begin{eqnarray}
A &=& \frac{L^2}{(1 - {\rm e}^2)\;\gamma\,\overline{M}\,M_A\,M_B} \,,
\label{appendixB_47}
\\
B &=& \frac{L^2}{\sqrt{1 - {\rm e}^2}\;\gamma\,\overline{M}\,M_A\,M_B} \,.
\label{appendixB_48}
\end{eqnarray}

\noindent
To solve the integral (\ref{appendixB_35}), one substitutes 
\begin{eqnarray}
r_{AB} - A = - A\;{\rm e}\;\cos E\,,
\label{eccentric_anomaly_5}
\end{eqnarray}

\noindent
where $E$ is called eccentric anomaly. Then, one obtains for the integral in Eq.~(\ref{appendixB_35}) the expression
\begin{eqnarray}
t &=& \left( \frac{A^3}{\gamma\;(M_A + M_B)}\right)^{1/2} \; \int d E\;(1 - {\rm e} \; \cos E)\,,
\label{appendixB_49}
\end{eqnarray}

\noindent
and the solution is given by
\begin{eqnarray}
t &=& \left( \frac{A^3}{\gamma\;(M_A + M_B)}\right)^{1/2}\;(E - {\rm e}\;\sin E) \;,
\label{appendixB_50}
\end{eqnarray}

\noindent
where the integration constant vanishes, i.e. the particle at $t=0$ is in periastron.
The Eqs.~(\ref{appendixB_46}) and (\ref{appendixB_50}) are the general solutions of two-body problem.
They can be rewritten as
\begin{eqnarray}
r_{AB} &=& A \left( 1 - {\rm e} \;\cos E \right) \,,
\label{appendixB_51}
\\
\nonumber\\
t &=& \left(\frac{A^3}{\gamma\,(M_A + M_B)}\right)^{1/2} \left( E - {\rm e} \; \sin E \right)\,.
\label{appendixB_52}
\end{eqnarray}

\noindent
In case of ellipse, $E = 0$ in periastron, $E = \pi$ in apastron, and for a complete orbit $E$ runs
from $E = 0$ to $E= 2\,\pi$. Thus, one obtains for the orbital period the expression
\begin{eqnarray}
T &=& 2 \,\pi \; \left( \frac{A^3}{\gamma\;(M_A + M_B)}\right)^{1/2}\;.
\label{appendixB_55}
\end{eqnarray}

\noindent
Note, the solution $\ve{r}$ in Cartesian coordinates,
$x = r_{AB}\;\cos \phi$ and $y = r_{AB}\;\sin \phi$:
\begin{eqnarray}
\ve{r}_{AB} &=& \left( \begin{array}[c]{l}
x \\
\displaystyle
y
\end{array}\right),
\label{appendixB_56}
\\
\nonumber\\
x &=& A \left( \cos E - {\rm e} \right) \,,
\label{appendixB_57}
\\
\nonumber\\
y &=& A \left( 1 - {\rm e}^2 \right)^{1/2} \;\sin E \,.
\label{appendixB_58}
\end{eqnarray}

\noindent
The coordinates of the bodies A and B, i.e. their orbits, are given by
\begin{eqnarray}
\ve{r}_A &=& \ve{r}_{\rm CMS} + \frac{\ve{r}_{AB}}{1 + \frac{\displaystyle M_A}{\displaystyle M_B}}\,,
\label{appendixB_61}
\\
\nonumber\\
\ve{r}_B &=& \ve{r}_{\rm CMS} - \frac{\ve{r}_{AB}}{1 + \frac{\displaystyle M_B}{\displaystyle M_A}}\,.
\label{appendixB_62}
\end{eqnarray}

\begin{figure}[h!]
\centering
\includegraphics[width=6.0cm,angle=270]{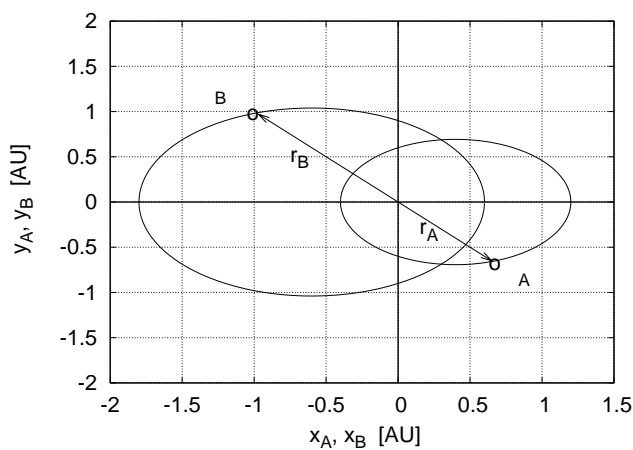}
\caption{Geometrical representation of the coordinates of a binary star.
In the example considered, the masses are $M_A=1.5\;M_{\odot}$ and $M_B=1.0\;M_{\odot}$,
respectively. The semi-major axis of the binary system is chosen ${\rm A}=2\;{\rm AU}$ and eccentricity
is taken ${\rm e}=0.5$. The coordinates of mass center are $\ve{r}_{\rm CMS} = \ve{0}$.
The massive bodies A and B are always in opposition to each other.}
\label{FIG: Ellipse}
\end{figure}

\noindent
Accordingly, the geometry of the orbit is determined by two orbital parameters: semi-major axis A and
eccentricity e. In order to know the position of one celestial body, either component A or component B,
two additional orbital parameters are needed, namely orbital period $T$ and true anomaly $\nu$.
A geometrical representation of the coordinates of the components of
a binary star is given in FIG.~\ref{FIG: Ellipse} for the case of
$M_A = 1.5\,M_{\odot}, M_B = 1.0\,M_{\odot}\,, {\rm e} = 0.5, A = 2\,{\rm AU}$.

\section{Derivation of Eq.~(\ref{inclination_22})\label{AppendixC}}

For the inclination formula the impact of eccentricity on light deflection is neglected, thus $e = 0$,
implying that $\omega = 0$ is taken. Then, for the vectors from massive body to observer $\ve{x}_1$
and from massive body to source $\ve{x}_0$, one has 
\begin{equation}
\ve{x}_1 = r\;\left( \begin{array}[c]{c}
\displaystyle
\sin i - \epsilon_1\,\cos E\\
\nonumber\\
\displaystyle
- \epsilon_1\,\sin E \\
\nonumber\\
\displaystyle
\cos i
\end{array}\right),
\label{appendixC_5}
\end{equation}

\vspace{0.5cm}

\begin{equation}
\ve{x}_0 =
- A\;\left( \begin{array}[c]{c}
\displaystyle
\cos E\\
\nonumber\\
\displaystyle
\sin E \\
\nonumber\\
\displaystyle
0
\end{array}\right),
\label{appendixC_10}
\end{equation}

\noindent
where the small parameter
\begin{eqnarray}
\epsilon_1 = \frac{\displaystyle A}{\displaystyle r}\,\frac{\displaystyle m_B}{\displaystyle m_A + m_B} \ll 1 
\label{epsilon_1}
\end{eqnarray}

\noindent
has been introduced. From Eqs.~(\ref{appendixC_5}) and (\ref{appendixC_10}) one obtains for vector $\ve{k} = \ve{R}/R$,
where $\ve{R} = \ve{x}_1 - \ve{x}_0$, the expression
\begin{equation}
\ve{k} =
\frac{1}{\sqrt{1 + 2\,\epsilon_2\,\sin i\,\cos E + \epsilon_2^2}}
\left( \begin{array}[c]{c}
\displaystyle
\sin i + \epsilon_2\,\cos E\\
\nonumber\\
\displaystyle
\epsilon_2\,\sin E \\
\nonumber\\
\displaystyle
\cos i
\end{array}\right),
\label{appendixC_15}
\end{equation}

\noindent
where the small parameter
\begin{eqnarray}
\epsilon_2 = \frac{\displaystyle A}{\displaystyle r}\,\frac{\displaystyle m_A}{\displaystyle m_A + m_B} \ll 1 
\label{epsilon_2}
\end{eqnarray}

\noindent
has been introduced. Note that (\ref{appendixC_10}) and (\ref{appendixC_15}) yields
\begin{eqnarray}
d &=& \left|\ve{k} \times \ve{x}_0 \right| = A\,\left|\cos i \right| \left( 1 + {\cal O} \left(\epsilon_2\right)\right).
\label{expression_d}
\end{eqnarray}

\noindent
Using (\ref{appendixC_5}) - (\ref{epsilon_2}), the generalized lens equation (\ref{generalized_lens_1}) reads
\begin{eqnarray}
&& \varphi = \frac{1}{2}\,\frac{1}{T_1\,T_2} 
\left(\sqrt{W_1^2 + 8 \,\frac{m}{r}\frac{A}{r} \left(T_0 + T_1 - \epsilon_1\right)T_2} - W_1\right),
\label{generalized_lens_2}
\end{eqnarray}

\noindent
where $\displaystyle W_1 = \frac{A}{r}\sqrt{1 - T_0^2}$ and 
\begin{eqnarray}
T_0 &=& \sin i \, \cos E\,,
\label{T_0}
\\
\nonumber\\
T_1 &=& \sqrt{1-2\,\epsilon_1\,\sin i\,\cos E + \epsilon_1^2}\,,
\label{T_1}
\\
\nonumber\\
T_2 &=& \sqrt{1+2\,\epsilon_2\,\sin i\,\cos E + \epsilon_2^2}\,.
\label{T_2}
\end{eqnarray}

\noindent
By series expansion one obtains up to terms of order 
$\displaystyle {\cal O} \left(\frac{A}{r}\,\sqrt{\frac{m}{r} \frac{A}{r}}\right)$:
\begin{eqnarray}
&& \varphi = \frac{1}{2} \left(\sqrt{W_2^2 + 8\, \frac{m}{r}\frac{A}{r}
\left(1 + w\right)} - W_2\right),
\label{generalized_lens_3}
\end{eqnarray}

\noindent
where $\displaystyle W_2 = \frac{A}{r} \sqrt{1 - w^2}$ with $w = \sin i\,\cos E$.
The minimal and maximal light deflection angle are
\begin{eqnarray}
\varphi_{\rm min} &=& \varphi \left(i = \frac{\pi}{2}\,, E=\pi\right) = 0\,,
\label{min}
\\
\nonumber\\
\varphi_{\rm max} &=& \varphi \left(i = \frac{\pi}{2}\,, E=0\right) = 2\,\frac{\sqrt{m\,A}}{r}\,.
\label{max}
\end{eqnarray}

\noindent
In this study, the maximal possible light deflection effect is of interest. Accordingly, two configurations are relevant: 
\begin{eqnarray}
&&\varphi \left(E=0\right) = \frac{1}{2} \left(\sqrt{\frac{A^2}{r^2} \, \cos^2 i  + 8 \, \frac{m}{r}
\frac{A}{r} \left(1 + \sin i\right)} - \frac{A}{r}\left| \cos i \right|\right),
\label{appendixC_24_C}
\end{eqnarray}

\noindent
up to order ${\cal O} \left(\frac{A}{r}\,\sqrt{\frac{m}{r}\,\frac{A}{r}}\right)$
which is just Eq.~(\ref{inclination_22}), and 
\begin{eqnarray}
&& \varphi \left(i=\frac{\pi}{2}\right) = \frac{1}{2} \left(\sqrt{\frac{A^2}{r^2} \, \sin^2 E  + 8\, \frac{m}{r}
\frac{A}{r} \left(1 + \cos E\right)} - \frac{A}{r} \left| \sin E \right|\right),
\label{appendixC_24_D}
\end{eqnarray}

\noindent
up to order $\displaystyle {\cal O} \left(\frac{A}{r}\,\sqrt{\frac{m}{r} \frac{A}{r}}\right)$.
Furthermore, it is useful to take into account only astrometric positions with
$\displaystyle 0 \le E \le \frac{\pi}{2}$, because otherwise the light deflection is for sure negligible.

\section{Derivation of Eq.~(\ref{inclination_30_A})\label{AppendixD}}

From (\ref{appendixC_24_C}) one obtains 
\begin{eqnarray}
\left( 2\,\varphi + \frac{A}{r}\,\left|\cos i \right| \right)^2 &=& \frac{A^2}{r^2}\,\cos^2 i
+ 8\,\frac{m}{r}\,\frac{A}{r}\,\left(1+\sin i\right).
\label{appendixD_10}
\end{eqnarray}

\noindent
From (\ref{appendixD_10}) one obtains 
\begin{eqnarray}
&& \left(\varphi^2 + 4\,\frac{m^2}{r^2}\right)\frac{A^2}{r^2}\sin^2 i + 4\frac{m}{r}\,\frac{A}{r}
\left(2\,\frac{m}{r}\frac{A}{r} - \varphi^2\right) \sin i 
\nonumber\\
&=&
\left(\frac{A^2}{r^2} + 4\,\frac{m}{r}\frac{A}{r}\right)\varphi^2
- 4\frac{m^2}{r^2}\,\frac{A^2}{r^2} - \varphi^4 .
\label{appendixD_15}
\end{eqnarray}

\noindent
Eq.~(\ref{appendixD_15}) represents an quadratic equation for the expression $\left|\,\sin i\,\right|$,
which has the following both solutions for the inclination $i$:
\begin{eqnarray}
\sin i &=& \left( - \frac{p}{2} \pm \sqrt{\frac{p^2}{4} - q }\right),
\label{appendixD_21}
\end{eqnarray}

\noindent
where
\begin{eqnarray}
p &=& \frac{8\,m^2\,A - 4\,m\,r^2\,\varphi^2}{A \left(r^2\,\varphi^2 + 4\,m^2\right)} \,,
\label{appendixD_30}
\\
\nonumber\\
q &=& - \,\frac{A^2\,r^2\,\varphi^2 + 4\,m\,A\,r^2\,\varphi^2-4\,m^2\,A^2 - r^4\,\varphi^4}
{A^2 \left(r^2\,\varphi^2 + 4\,m^2\right)}\,.
\label{appendixD_35}
\end{eqnarray}

\noindent
Eq.~(\ref{appendixD_21}) represents two solutions, however only the one with the plus-sign is valid.
This can be shown as follows. For the value $i = \frac{\displaystyle \pi}{\displaystyle 2}$ the light deflection
has to be $\varphi = \varphi_{\rm max} = 2\sqrt{m\,A}/r$, according to (\ref{inclination_25}). By inserting
$\varphi_{\rm max}$ in Eqs.~(\ref{appendixD_30}) and (\ref{appendixD_35}) one obtains $p=-2\,m/(A + m)$ and
$q = - (A - m)/(A + m)$. If one inserts $i = \frac{\displaystyle \pi}{\displaystyle 2}$ for $p$ and $q$ into
Eq.~(\ref{appendixD_21}) one obtains the relation
\begin{eqnarray}
1 &=& \frac{m}{A + m} \pm \sqrt{\frac{m^2}{\left(A + m \right)^2} + \frac{A - m}{A + m}}
= \frac{m}{A + m} \pm \frac{A}{A + m}\,.
\label{appendixD_40}
\end{eqnarray}

\noindent
Obviously, relation (\ref{appendixD_40}) is only correct for the upper sign.
A very similar proof can also be done using $\varphi_{\rm min}$ which also yields
that the upper sign is the correct and only solution. Therefore, the inclination formula
is given by (note, that in the region under consideration $\sin i = \sin \left(\pi - i\right)$)
\begin{eqnarray}
i =
\begin{array}[c]{c}
\displaystyle
\arcsin \left( - \frac{p}{2} + \sqrt{\frac{p^2}{4} - q }\right)  \quad {\rm for} \quad 0 \le i \le \frac{\pi}{2} \,,
\\
\\
\displaystyle
\pi - \arcsin \left( - \frac{p}{2} + \sqrt{\frac{p^2}{4} - q }\right) \quad {\rm for} \quad \frac{\pi}{2} < i \le \pi\,.
\end{array}
\label{appendixD_45}
\end{eqnarray}

\noindent
For the complete region $0 \le i \le \pi$ one obtains for the inclination formula the following expression:
\begin{eqnarray}
\left|\,\frac{\pi}{2} - i\,\right| &=& \arccos \left(- \frac{p}{2} + \sqrt{\frac{p^2}{4} - q } \right),
\label{inclination_55}
\end{eqnarray}

\noindent
where $p$ and $q$ are given by Eqs.~(\ref{appendixD_30}) and (\ref{appendixD_35}).

\section{Derivation of Eq.~(\ref{KMS_D})}\label{Appendix_KMS}

In this appendix, some basic steps of the calculations of 
{\it Klioner, et al.} \cite{inclination_formula} are presented. 
A scheme of the light propagation in a binary system is 
presented by Fig.~\ref{FIG: Distance1}. The light signal from component B, considered to be the light source, is deflected by 
component A, considered to be the massive body which curves the space-time. 
The vector $\ve{x}_1$ points from the mass center of massive body to the observer, and vector $\ve{x}_0$
points from the mass center of massive body to the source;
$\ve{R} = \ve{x}_1 - \ve{x}_0$ and its absolute value $R = \left|\ve{R}\right|$, and
$\displaystyle m = \frac{G\,M}{c^2}$ is the Schwarzschild radius of massive body; the explicit label A is omitted. 
Furthermore we define the impact vector $\ve{d}=\ve{k} \times \left( \ve{x}_1 \times \ve{k} \right)$ is defined, 
and its absolute value $d=\left|\ve{d}\right|$, see also FIG.~\ref{FIG: Distance1}. 

According to {\it Klioner {\rm \&} Zschocke} \cite{Article_Klioner_Zschocke}, the transformation of $\ve{k}$ to the unit 
tangent vector $\ve{n}$ of light-trajectory at observer is in standard post-Newtonian approach given by
\begin{eqnarray}
\ve{n} &=& \ve {k} - 2\,m\,\frac{\ve{k} \times \left(\ve{x}_0 \times \ve{x}_1\right)}
{x_1 \left(x_0\,x_1 + \ve{x}_0\cdot\ve{x}_1\right)} + {\cal O} \left(m^2\right).
\label{Appendix_KMS_A}
\end{eqnarray}

\noindent
Note, the PPN parameter of parameterized post-Newtonian formalism, which characterizes a possible deviation of the physical 
reality from general theory of gravity, is set equal to $1$ here for simplicity. 
This expression is valid as long as $d \gg m$, but diverges for $d \rightarrow 0$. Thus, it is not valid for all 
possible binary configurations, instead one has take care to consider only such astrometric configurations with $d \gg m$. 
By means of (\ref{Appendix_KMS_A}) one obtains for the light deflection angle $\varphi$, i.e. for the 
angle between $\ve{n}$ and $\ve{k}$, the expression
\begin{eqnarray}
\varphi &=& 2\,\frac{m}{r}\,\tan \frac{\psi}{2}\,,
\label{Appendix_KMS_B}
\end{eqnarray}

\noindent
where $\frac{\displaystyle \sin \psi}{\displaystyle 1 + \cos \psi}
= \tan \frac{\displaystyle \psi}{\displaystyle 2}$, $x_1 = r + {\cal O} \left( A \right)$ has been used, and $\psi$ is the 
angle between $\ve{x}_0$ and $\ve{x}_1$. 
The expression (\ref{Appendix_KMS_B}) diverges for $\psi \rightarrow \pi$, which 
corresponds with the mentioned divergence of (\ref{Appendix_KMS_A}) for $d \rightarrow 0$.
Obviously, $\psi \le i + \frac{\displaystyle \pi}{\displaystyle 2}$
(from Eq.~(\ref{observable_30}) it is clear that eccentric anomaly $E$ of binary system should be very close to
zero for the light deflection effect to be observable at the level of microarcsecond, i.e. one 
actually could even assume $\psi \simeq i + \frac{\displaystyle \pi}{\displaystyle 2}$) and one obtains
\begin{eqnarray}
\varphi &\le& 2\,\frac{m}{r}\,\tan \left(\frac{i}{2} + \frac{\pi}{4}\right)
= 2\,\,\frac{m}{r}\, \cot \left(\frac{\pi}{4} - \frac{i}{2}\right),
\label{Appendix_KMS_C}
\end{eqnarray}

\noindent
where $\tan \left(\alpha + \frac{\displaystyle \pi}{\displaystyle 2}\right) = - \cot \alpha$ has been used,
$\cot \alpha = \tan^{-1} \alpha$, and the asymmetry of function $\cot \alpha$. From (\ref{Appendix_KMS_C}) one obtains 
\begin{eqnarray}
\left|\frac{\pi}{2} - i \right|_{\rm KMS} &\le& 2\,\arctan \left(0.0197\;\frac{M}{M_{\odot}}\;
\frac{\mu{\rm as}}{\varphi}\;\frac{\rm pc}{r}\right),
\label{Appendix_KMS_D}
\end{eqnarray}

\noindent
where $\displaystyle \frac{m}{m_{\odot}}=\frac{M}{M_{\odot}}$ has been used (recall $M$ is the mass of component A and 
$M_{\odot}$ is the solar mass), and the numerical values 
$m_{\odot} \simeq 1.476 \times 10^3\,{\rm m}$, 
$1\,\mu{\rm as} \simeq 4.848 \times 10^{-12}\,{\rm rad}$ and $1\,{\rm pc} \simeq 3.086 \times 10^{16}\,{\rm m}$ 
have been inserted, so that $\displaystyle \frac{2\,m_{\odot}}{\mu {\rm as}\,{\rm pc}} \simeq 0.0197$. The simplified 
inclination formula (\ref{Appendix_KMS_D}) has first been obtained by {\it Klioner, et al.} 
\cite{inclination_formula}; here the fact is noticed, that due to the divergence of the post-Newtonian solution 
(\ref{Appendix_KMS_A}) for $d\rightarrow 0$, which corresponds to 
$\psi \rightarrow \pi$, the applicability of (\ref{Appendix_KMS_D})
is restricted by the condition $d \gg m$. Using $d = A\,\left|\cos i\right|$ one obtains the validity condition for
the applicability of (\ref{Appendix_KMS_D}): 
\begin{eqnarray}
\left|\frac{\pi}{2} - i \right|_{\rm KMS} \gg \arcsin \frac{m}{A}\;.
\label{Appendix_KMS_E}
\end{eqnarray}

\noindent
Equation (\ref{Appendix_KMS_D}) provides a relation between inclination $i$ and light
deflection $\varphi$ for a binary system characerized by it's distance
$r$ from the observer and it's stellar mass $M$; note that (\ref{Appendix_KMS_D}) agrees
with the simplified inclination formula given by Equation (\ref{inclination_50}). 

\section{Probability Distribution}\label{Probability_Distribution}

Assume a probability distribution of a quantity $x$ is given by $f(x)$.
The probability $P$, to find a system in the interval $x_i \le x \le x_i + \Delta x$ is given by
\begin{eqnarray}
P (x_i \le x \le x_i + \Delta x) &=& \frac{\displaystyle
\int\limits_{x_i}^{x_i + \Delta x} d z\;f(z)}{
\displaystyle\int\limits_{x_{\rm min}}^{x_{\rm max}}
d z\;f(z)}\,,
\label{probability_distribution_5}
\end{eqnarray}

\noindent
where the region of validity of probability distribution $f(x)$ is given by $x_{\rm min}$ and $x_{\rm max}$.
In the infinitesimal limit $\Delta x \rightarrow d x$, one obtains by series expansion
the following explicit form for the here used probability distributions: for a power law
$f(x) \sim x^{-\alpha}$ with $\alpha \neq 1$ one finds 
\begin{eqnarray}
f (x) &=& \frac{\displaystyle \left(1-\alpha\right)\;x^{-\alpha}}{
\displaystyle x_{\rm max}^{\left(1-\alpha\right)} - x_{\rm min}^{\left(1-\alpha\right)}}\,,
\label{probability_distribution_10}
\end{eqnarray}

\noindent
and for a logarithmic law $f(x) \sim x^{-1}$ one has 
\begin{eqnarray}
f(x) &=& \frac{1}{x}\;\left(\ln \frac{x_{\rm max}}{x_{\rm min}}\right)^{-1}\,.
\label{probability_distribution_15}
\end{eqnarray}

\noindent
The normalization is $\int\limits_{x_{\rm min}}^{x_{\rm max}} f(x)\,dx= 1$
and the averaged value $\overline{x} = \int\limits_{x_{\rm min}}^{x_{\rm max}} f(x)\,x\,dx$.

\end{appendix}

\end{document}